\documentclass[graybox]{svmult}

\usepackage{graphicx}        % standard LaTeX graphics tool
\def\msol   {\ifmmode{{\rm M}_{\odot}}\else{M$_{\odot}$}\fi}

\begin{document}

\title*{Explaining the formation of bulges with MOND}
\titlerunning{Bulges in MOND} 
% your contribution title if the original one is too long
\author{Fran\c{c}oise Combes}
% Use \authorrunning{Short Title} for an abbreviated version of
% your contribution title if the original one is too long
\institute{Observatoire de Paris, LERMA, CNRS, 61 Av. de l'Observatoire, F-75014, Paris, France \email{francoise.combes@obspm.fr}}
%
% Use the package "url.sty" to avoid
% problems with special characters
% used in your e-mail or web address
%

\maketitle

\abstract{In the cold dark matter (CDM) paradigm, bulges easily form through
galaxy mergers, either major or minor, or through clumpy disks in the early universe,
where clumps are driven to the center by dynamical friction. Also pseudo-bulges,
with a more disky morphology and kinematics, can form more slowly through 
secular evolution of a bar, where resonant stars are elevated out of the plane,
in a peanut/box shape. As a result, in CDM cosmological simulations, 
it is very difficult to find a bulgeless galaxy, while they are observed
very frequently in the local universe. 
A different picture emerges in alternative models of the missing
mass problem. In MOND (MOdified Newtonian Dynamics), galaxy mergers are much less
frequent, since the absence of dark matter halos reduces the dynamical friction
between two galaxies. Also, while clumpy galaxies lead to rapid classical bulge 
formation in CDM, the inefficient dynamical friction with MOND
in the early-universe galaxies prevents the clumps to coalesce together
in the center to form spheroids. This leads to less frequent and less massive classical 
bulges. Bars in MOND are more frequent and stronger, and have a more constant pattern speed,
which modifies significantly the pseudo-bulge morphology. The fraction 
of pseudo-bulges is expected to be dominant in MOND.}

\section{Introduction}
\label{sec:1}

Although the standard CDM model for dark matter is the best frame to represent
the universe at large scales, and account for galaxy formation, it experiences
difficulties at galaxy scale (e.g. Moore et al. 1999, Silk \& Mamon 2012).  
Cosmological simulations in the
standard model predict an over concentration of dark matter in galaxies, 
and cuspy density profiles, instead of the density cores derived from rotation curves,
especially in low-mass galaxies (e.g. de Blok et al 2008, Swaters et al 2009). 
Also simulations have difficulties to form large galaxy disks, since the angular 
momentum of baryons is lost against massive dark halos
(e.g. Navarro \& Steinmetz 2000), and the missing satellites problem remains
unsolved (Diemand et al. 2008). In addition, observed low-mass satellites
of the Milky Way have a much larger baryonic fraction than expected 
from halo abundance matching (e.g. Boylan-Kolchin et al. 2011, 2012).

A large numerical effort has been spent to solve these problems by 
the detailed physics of the baryonic component, in particular star formation
and AGN feedback (e.g. Vogelsberger et al. 2014, Schaye et al. 2015).
Another track is to explore alternatives to dark matter models, and in
particular modified gravity scenarios, able to account for the missing mass
in galaxies.

\begin{figure}[ht]
%\sidecaption
\centerline{
\includegraphics[width=11.0cm]{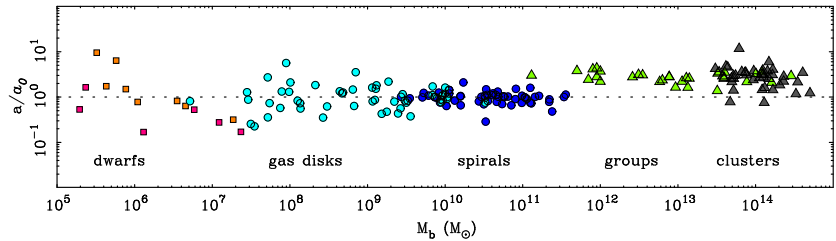}}
\caption{The observed parameter V$_{f}^4$/(GM$_b$),
where V$_{f}$ is the rotational velocity taken in the flat
portion of the rotation curve, and M$_b$ is the baryonic mass, can be also written as
the acceleration of the system: a = V$_{f}^2$/R, and R = GM /V$_{f}^2$.
The very small deviation of a from the constant a$_0$
is remarkable, given the large range of ten decades
in baryonic mass M$_b$. This observation is somewhat puzzling for the
standard dark matter model, but is the basis of the 
modified gravity (MOND) model (from Famaey \& McGaugh 2012).
}
 \label{fig1}
\end{figure}

Already 30 years ago, Milgrom (1983) had the idea
of the MOdified Newtonian Dynamics (MOND), based on the fundamental observation that the
missing mass problem occurs only in the weak field regime, at low acceleration,
when a is lower than the characteristic value of a$_0$= 2 10$^{-10}$m/s$^2$.
The observed flat rotation curves in the outer parts of galaxies 
suggests that in this regime the actual acceleration varies in 1/r. 
Galaxies are also following the baryonic Tully-Fisher relation 
(McGaugh et al. 2000), where the baryonic mass of a system is proportional to 
the 4th power of the maximum rotational velocity (see Figure 1). 
Milgrom then proposes that at acceleration below 
a$_0$= 2 10$^{-10}$m/s$^2$, the gravitational attraction will
tend to the formulation a = (a$_0$ a$_N$)$^{1/2}$, where a$_N$ is the Newtonian value. 
This effectively produces an acceleration in
1/r, implying a flat rotation curve in the limiting regime, and leading automatically 
to the Tully-Fisher relation. The transition
between the Newtonian and MOND regime is controlled by an interpolation function 
$\mu$(x), of x=a/a$_0$, which
standard form is $\mu$(x) = x /(1+x$^2$)$^{1/2}$. It essentially tends to x in 
the MOND regime, when x is smaller than 1, and to
unity in the Newtonian regime. This phenomenology has a large success explaining rotation curves and kinematics
of galaxies, from dwarf irregulars dominated by dark matter (and therefore in the 
MOND regime), to the giant spirals and
ellipticals, dominated by baryons (e.g. Sanders \& McGaugh 2002).
Although the model is still empirical, it is possible to build
relativistically covariant theories, able to reproduce gravitational
lensing and other phenomena, while tending asymptotically to
the above formulation in the non-relativistic limit (Bekenstein 2004).

\bigskip
The galaxy dynamics is quite different in the MOND hypothesis with respect to the 
standard dark matter model. Some phenomena have already been explored (see e.g. 
the review by Famaey \& McGaugh 2012), but many are still to be discovered, and
in particular galaxy formation, and high redshift evolution. The stability of
galaxy disks is fundamentally different, provided that they have low surface
brightness (LSB), and are close to the MOND regime (Milgrom \& Sanders 2007). Since the 
MOND disks are completely self-gravitating, they could be much more unstable, however
the acceleration is varying asymptotically as the square root of the mass
(and not linearly with the mass), so the final effects are not intuitive.
Bars are forming quickly in MOND disks, and their pattern speed is not declining
through dynamical friciton against a dark matter halo, so resonances are long-lived,
and may have more impact (Tiret \& Combes 2007). Galaxy interactions with no 
extended dark halos suffer much less dynamical friction, and mergers are rare 
(Tiret \& Combes 2008b). This changes very significantly the hierachical scenario
of galaxy formation, and in particular bulge formation.
Therefore, although bulges are now generally in the Newtonian regime today,
their formation is certainly very different in the MOND frame with respect to
the standard model.
Bulges are increasingly important along the Hubble sequence towards
the early-types, which correspond to the more massive end. For giant galaxies,
the low acceleration regime is encountered only in the outer parts, and the
central parts remain Newtonian. Only dwarf galaxies and LSB objects
without bulges are still in the MOND regime in their center. This means that
bulges today are not likely to be affected by a modified dynamics.

In the following, we will consider in turn the main dynamical mechanisms
to form bulges in the $\Lambda$CDM paradigm: 
\begin{itemize}
\item Mergers, major or a series of minor mergers
\item Secular evolution, bars and the formation of pseudo-bulges
\item Clumpy galaxies at high redshift and dynamical friction
\end{itemize}

Are all these processes also at work in MOND, and with which efficiency?
It is well known that the standard $\Lambda$CDM  model has difficulties
to account for the large number of observed bulge-less galaxies (Kormendy et al. 2010).
Is this problem solved by MOND?

\section{Galaxy mergers}
\label{sec:2}
In the standard hierarchical scenario, galaxy mergers play a large role
in mass assembly, and one of the results of the repeated coalescence of 
galaxies is to randomly average out the angular momentum of the system,
and to form spheroids (e.g. Toomre 1977, Barnes \& Hernquist 1991,
Naab \& Burkert 2003, Bournaud et al. 2005, 2007a). In these last works, it was
shown how repeated minor mergers progressively accumulate stars in 
a central spheroid and grow the bulge, to transform the galaxy in a more
early-type spiral. Eventually, N minor mergers of mass ratio N:1 result in 
an elliptical remnant quite similar to those formed in a 1:1 merger.
 As shown by Barnes (1988), mergers are very efficient in forming long tidal tails
while the main baryonic components merge quickly, because of the existence of
extended and massive dark halos, which take the orbital angular momentum away.
 It can then be expected that the frequency of mergers will depend crucially
on the model assumed for the missing mass.

\subsection{Major mergers in MOND}
\label{major-merger}

One of the main questions is to know whether the MOND dynamics is able to 
produce long tails in major mergers of galaxies, like in the prototypical
Antennae system (Figure 2). These tails have also helped to constrain
the dark matter halos potential (Dubinski et al. 1996). With MOND,
the result is not easy to predict, and numerical
simulations are necessary, since the External Field Effect (EFE) perturbs
the MONDian dynamics in the outer parts of galaxies.
This new effect particular to MOND comes from the fact that it violates the 
Strong Equivalence Principle of General Relativity. In the Newtonian frame,
the internal gravitational forces of a system are independent of their external
environment: if the object is embedded in a large system, exerting a force
which can be considered constant all over the object, then the internal dynamics
is unchanged. Of course, if the force is varying across the object,
its differential gives rise to tidal forces, which impact the object.
But in the MOND dynamics, even a constant force may create an acceleration 
above the critical one a$_0$, and get the object out of the MOND regime
(Milgrom 1983, 1998).

Several cases can be distinguished to model the EFE, according
to the respective values of the external acceleration a$_e$ with respect
to the internal acceleration a of the object under consideration, and the
critical acceleration a$_0$. 
If a$_e <$ a $<$ a$_0$, then the standard MOND effects are retrieved, 
and if a $<$ a$_0 <$ a$_e$, then the EFE is strong enough to make the
system purely Newtonian. But in the intermediate regime,
where  a $<$ a$_e <$ a$_0$ then the system is Newtonian with a renormalised
gravitational constant G. It can be estimated for instance in
a one-dimensional system, that the effective
gravitational constant is then G$_{eff}$ = G [$\mu_e$(1 + L$_e$)]$^{-1}$
where $\mu_e$ =$\mu$(a$_e$/a$_0$) and L$_e$ is the logarithmic gradient
of $\mu$ (Famaey \& McGaugh 2012). 
In the outer parts of a given system, the internal acceleration
is always vanishing, and there will always be a small a$_e <$ a$_0$, therefore
this represents the general case: the 
gravitational force falls again as 1/r$^2$, and the potential 
as 1/r and not logarithmically, as could be extrapolated. This allows to define
the escape velocity of the system, as in the Newtonian case. Computations of the EFE
in the Milky Way, due to the nearby Andromeda galaxy, have given results
compatible with the observations (Wu et al. 2007).

\begin{figure}[ht]
%\sidecaption
\centerline{
\includegraphics[width=11.0cm]{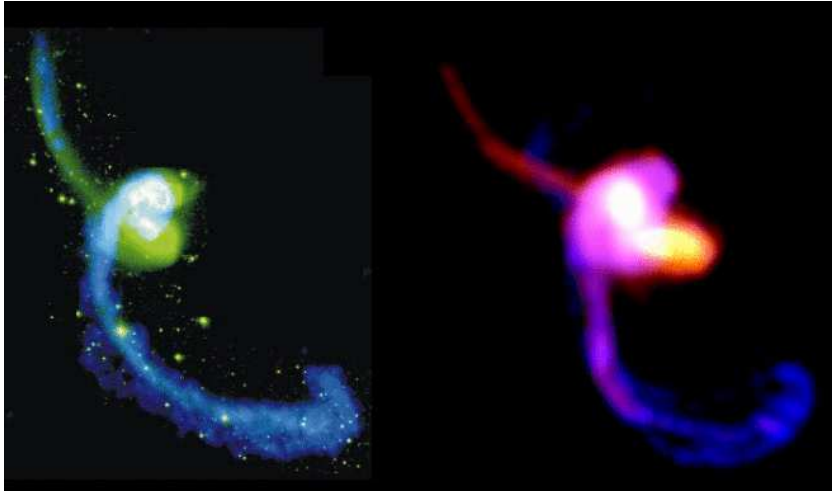}}
\caption{Simulation of tidal interactions in a major merger in MOND
(right, with gas in blue and stars in red) compared to the Antennae 
galaxies (Hibbard et al. 2001, HI gas in blue, and stars in green). The two
long tidal tails are reproduced (Tiret \& Combes 2008b).}
\label{fig2}
\end{figure}

Simulations with a 3D adaptive-mesh code able to solve the MOND equations,
 and including gas and stars, have shown that two long tidal tails can
develop in a major merger similar to the Antennae (cf Figure 2). In absence
of dark matter particles as receivers of the orbital angular momentum of
the two galaxies, baryons are playing this role, and tidal tails can be very
long. In addition, tidal dwarf galaxies can be naturally formed at the tip
of the tidal tails in MOND, while it requires radially extended dark matter halos
 in the standard model (Bournaud et al. 2003).
  The big difference between the two models is the efficiency of the dynamical friction.
While mergers can take only one orbit, or less than one Gyr in the standard model,
it will take several Gyrs with MOND, and mergers will occur only with selected
impact parameters, and initial relative angular momenta. 
At a distance of $\sim$ 100~kpc, two galaxies in circular orbits will not
merge in a Hubble time with MOND (Figure 3). On the contrary, in the standard model, 
galaxies have already plunged well inside their dark matter halos, of
radius $\sim$ 200~kpc. Then local dynamical friction is already effective, 
while in the MOND case, the relative decay relies only on the friction at distance,
which is much weaker.

\subsection{Dynamical friction}
\label{dyn-fric}

As described above, the gravitational forces between galaxies at large distance
are likely to vary as 1/r$^2$ as in the Newtonian regime, but with a boosted
constant, so the long-distance approach of galaxies could be thought similar. 
However the phenomena associated to dynamical friction are completely different.
Answers to this problem have been controversial at the start, since
Ciotti \& Binney (2004) computed the relaxation time in the MOND regime
with strong approximations: very small fluctuations, impulse approximation
for deflection or orbits, linear summation of effects, etc. and they compare
this two body relaxation time with that
in the Newtonian regime, considering the dark matter halo as a rigid
background, not participating in the fluctuations. Then they extrapolate their
finding of a shorter relaxation time in MOND to the dynamical friction
time, obtained for test particles for the local formula of Chandrasekhar (1943),
and conclude that globular clusters should spiral inwards 
to the center in dwarf galaxies in a few dynamical time-scales,
as well as galaxies in groups and in clusters. Nipoti et al. (2008) tried
to confirm these findings in simulations, by applying the same hypotheses of 
a tiny perturbation:
the massive bodies subject to the friction, either globular clusters or a 
rigid bar, have to contain less than 5\% of the baryonic mass, so that particles
absorbing the energy and angular momentum are not globally perturbed.
In realistic systems though, Nipoti et al. (2007) found that the merging timescales 
 for spherical systems are significantly longer in MOND than in Newtonian gravity 
with dark matter, and Tiret \& Combes (2007) found that bars keep their pattern speed
constant in MOND, while they are strongly slowed down in the Newtonian equivalent
system with a dark matter halo.
In summary, dynamical friction is very slow in MOND, since
galaxies are not embedded in extended and massive spheroids of dark matter particles, 
able to accept the orbital angular momentum. A short
merging time-scale for equal-mass interacting galaxies, 
as short as the CDM, is possible only for nearly radial orbits.
Although the impact of very small fluctuations could be larger in MOND
than in Newtonian dynamics, the effect saturates quickly when the perturbation
is no longer infinitesimal, and on the contrary the equivalent Newtonian system
with dark matter has shorter response time-scales, and 
a massive body (either a companion, or a galactic bar) is slowed down very efficiently.
The case of bars, their pattern speeds, and their impact on bulge formation
 will be detailed in the next Section \ref{sec:3}.

\begin{figure}[ht]
%\sidecaption
\centerline{
\includegraphics[width=11.0cm]{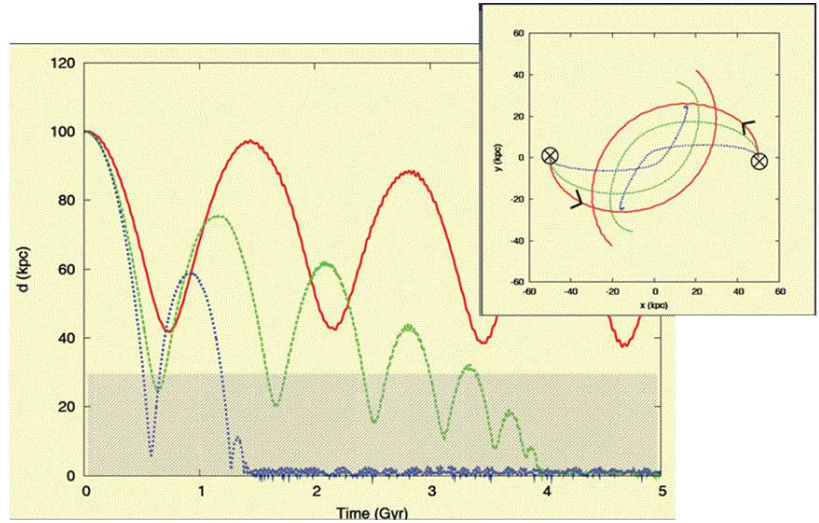}}
\caption{Radial decay during the tidal interactions between two
equal-mass spiral galaxies, in the MOND model (Tiret \& Combes 2008b).
 At left is shown the relative distance in kpc versus time in Gyr, while 
the insert at right shows the corresponding trajectories.
(image by O. Tiret).}
\label{fig3}
\end{figure}

\bigskip
Could the smaller merger frequency predicted by MOND be tested in observations?
Unfortunately, the actual merger frequency is not directly accessible.
 Observers tend to quote galaxy pair frequency, or starbursts due to 
mergers (e.g. Bell et al. 2006, Lopez-Sanjuan et al. 2013, Stott et al. 2013).
However, there is a degeneracy here, since galaxy can appear in pairs 
during either a short or long time-scale, and starbursts can occur at each 
closer passage.
In the standard DM models, an assumption is done on the duration of 
galaxy interactions, and the number of starbursts: according to the
initial relative velocity and the geometry of the encounter, the merger
is expected to occur in one or two passages. An intense starburst is associated to 
the final phase, and the number of starbursts is thought to count the number of 
mergers (e.g. Di Matteo et al 2007). In the MOND model, many
passages in binary galaxies will be required before the final merging, and 
a starburst may be triggered at each
pericenter. The number of starbursts as a function of redshift could then be similar, 
and cannot discriminate the two models.  The degeneracy cannot be raised between
a limited number of long-lived mergers, or a high frequency of short-lived mergers.

\section{Bars}
\label{sec:3}

To probe realistically the stability of disks with the MOND dynamics,
numerical simulations have been run, solving the N-body problem on a
grid, through the equations of Bekenstein \& Milgrom (1984).
 Brada \& Milgrom (1999) showed that disks were always more instable in MOND.
 For the equivalent Newtonian system with a spherical dark halo, the more unstable
galaxies are those with massive disks, which are more self-gravitating, while low-mass
disks are stabilized by their halo. In MOND, the instability is about the same 
for massive disks, which  are still in the Newtonian regime, however, low-mass disks
remain unstable, and their growth rate tend to a constant, instead of vanishing.

\subsection{Disk stability in MOND}
\label{stab}
From detailed comparison of two identical initial disks simulated
with Newtonian dynamics+dark matter and MOND,
 Tiret \& Combes (2007) have shown that bars develop quicker with modified 
gravity (see Figure 4). To have identical starts, the baryonic disk is first
computed in equilibrium with its velocity distribution in MOND, and then,
the amount of dark matter required to obtain the same derived rotation curve, 
is added for the Newtonian dynamics run.
 The evolution of the bar strength in Figure 4 reveals that both bars
experience a drop in their strength, and this is due to the vertical resonance,
building a peanut-shape feature, evolving in a  pseudo-bulge (e.g.
Combes \& Sanders 1981, Combes et al. 1990, Bureau \& Freeman 1999). The peanut occurs
later in MOND. The bar remains strong during a longer time-scale, but then
weakens, while the Newtonian bar can strengthen again, by exchanging
angular momentum with the dark halo (e.g. Athanassoula 2002).

\begin{figure}[ht]
%\sidecaption
\centerline{
\includegraphics[width=6.0cm]{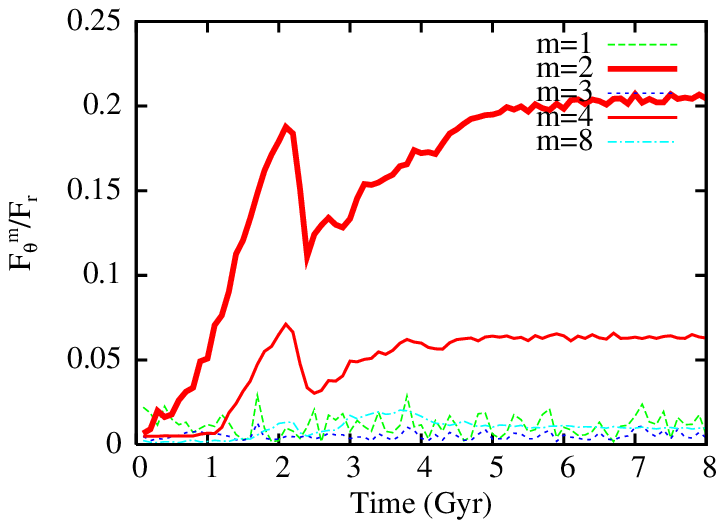}
\includegraphics[width=6.0cm]{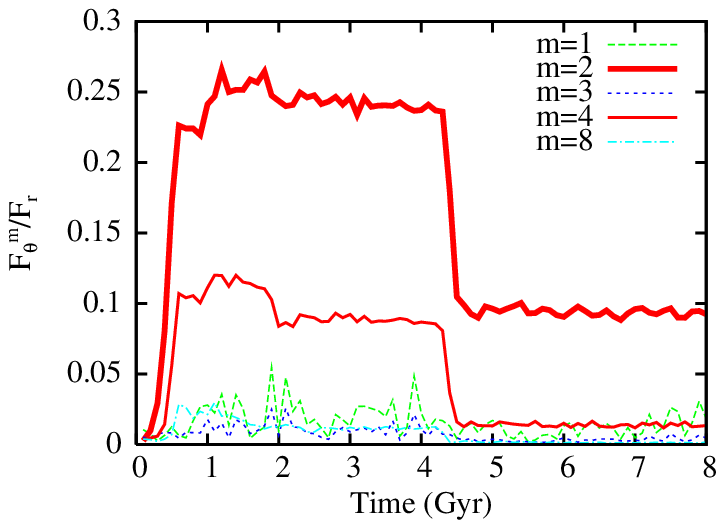}}
\caption{Strength of the bar formed in an Sa-type galaxy purely stellar simulation, 
measured by its Fourier harmonics m=2,3,4 and 8 (ratio of tangential to radial force),
for the CDM-Newton model (left) and MOND (right). The bar settles earlier in MOND, 
and stays longer, but after dropping at 4.5 Gyr, it does not develop again
as in the CDM (cf Tiret \& Combes 2007). The drop at 2.5 Gyr in the DM model
as in the MOND model at 4.5 Gyr is due to the formation of a peanut bulge,
through the vertical resonance (e.g. Combes et al. 1990). }
\label{fig4}
\end{figure}

This different way of growing results also in a different final morphology of the
stellar disks: in MOND the disk is more extended, since the bar has grown by 
angular momentum exchange with the outer disk particules. 
Figure 4 represents an early-type spiral Sa. When all types are considered,
the bar occurs much later in Newtonian models, because later types
are more dominated by the dark matter halo, and are less self-gravitating. In 
MOND it is the contrary, the bar is first stronger in late-types, and then the disk 
is heated too much and the bar weakens. When the statistics are computed
over the whole Hubble sequence, it appears that bars are stronger and more frequent in
MOND, when only stellar components are taken into account.
The higher MOND bar frequency is more in agreement with observations,
where 2/3rds of spiral galaxies are barred (e.g. Laurikainen
et al. 2004, 2009).

\subsection{Pattern speed evolution}
\label{pspeed}

The bar pattern speed evolutions are also different in the two models.
As shown in Figure 5 left, $\Omega_{bar}$ is almost constant in MOND,
while it drops by a factor 3 in 7 Gyr time in the equivalent Newtonian
system. This is clearly due to the exchange of angular momentum from
the bar to the dark matter halo, through dynamical friction. Indeed,
the test run when the Newtonian system is computed with a rigid halo, 
which cannot deform and produce dynamical friction, has an almost
constant  $\Omega_{bar}$ too.  

\begin{figure}[ht]
%\sidecaption
\centerline{
\begin{tabular}{lr}
\includegraphics[width=6.0cm]{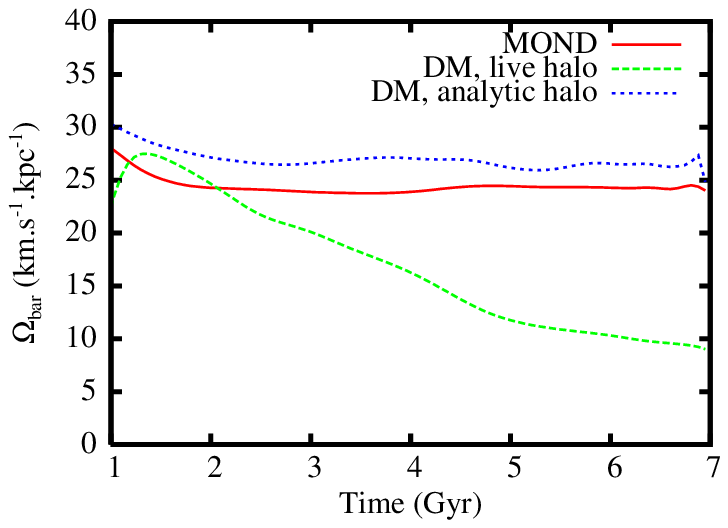} &
\shortstack{
\includegraphics[width=6.0cm]{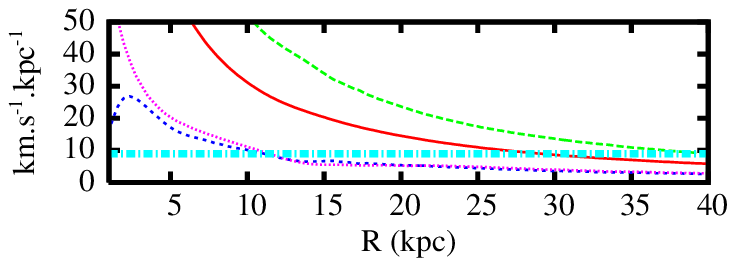}\\
\includegraphics[width=6.0cm]{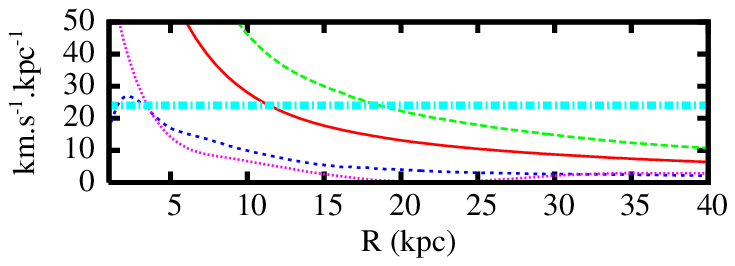}} \\
\end{tabular}}
\caption{{\bf Left} Bar pattern speeds versus time: in MOND, the pattern speed
remains constant, as in the Newtonian galaxy with 
a rigid dark matter halo. When the dark
halo particles are taken into account self-consistently, the bar slows down, losing
its angular momentum through dynamical friction.
{\bf Right} Frequency curves (from bottom to top, $\Omega-\kappa/2$, $\Omega-\nu_z/2$,
$\Omega$ and $\Omega+\kappa/2$) for the CDM case (top) and MOND (bottom).
The thick horizontal line is the pattern speed of the bar in each case 
(cf Tiret \& Combes 2007). }
\label{fig5}
\end{figure}

This drop in  $\Omega_{bar}$ for the Newtonian+dark matter model
has several consequences: First the Lindblad resonances in the plane
and the vertical resonance move in radius, as shown in Figure 5 right.
The pattern speed at the end of the simulation is shown as a thick dash line,
and the inner/vertical resonance moves from 2 kpc to 12 kpc.  Since the peanut
represents stars vertically up-lifted at resonance, this means that the radius of 
the peanut is moving radially outwards, as shown in Figure 6. In MOND on the contrary,
resonances are more long-lived, and can produce more robust effects.

\subsection{Bulges and pseudo-bulges}
\label{pseudo}
Until now, the comparison between MOND and the Newtonian equivalent systems
has been discussed with purely stellar disks.
However, the presence of gas, and its interaction with stars change the picture.
 Gas as a dissipational component, is subject to a phase shift in its response
to the bar pattern. There is a torque from the bar to the gas, that drives
it to the center. This changes the potential there, and therefore the $\Omega$ frequencies
and the resonances. The final result is a weakening of the bar, which can only
develop again through gas accretion (e.g. Bournaud \& Combes 2002).
Gas dissipation and star formation have been taken into account in MOND 
simulations by Tiret \& Combes (2008a).
Statistically, bars occur even more rapidly in gas rich disks, and especially in the 
Newtonian models, which were too stable in the purely stellar disks. This
makes the two models more similar, as far as the frequency of bars is concerned.
Since the baryonic mass is more concentrated with gas in any model, the vertical
resonance and the peanut occur at smaller radii, therefore the pseudo-bulges are
smaller and more boxy in appearance.

Finally, the gas is driven by gravity torques inwards inside corotation,
and outwards outside. It accumulates in rings at the inner (outer) Lindblad
resonances respectively, in star forming rings that reproduce the blue rings
observed in barred galaxies (e.g. Buta \& Combes, 1996). In MOND,
this phenomenon is even more remarkable, since first bars are still 
stronger and more frequent than in the Newtonian dynamics, but also the
exchange of angular momentum between the stellar and the gas components
is favored, while in the Newtonian case, there is competition with the dark halo
for this exchange.

\begin{figure}[ht]
%\sidecaption
\centerline{
\includegraphics[width=6.0cm]{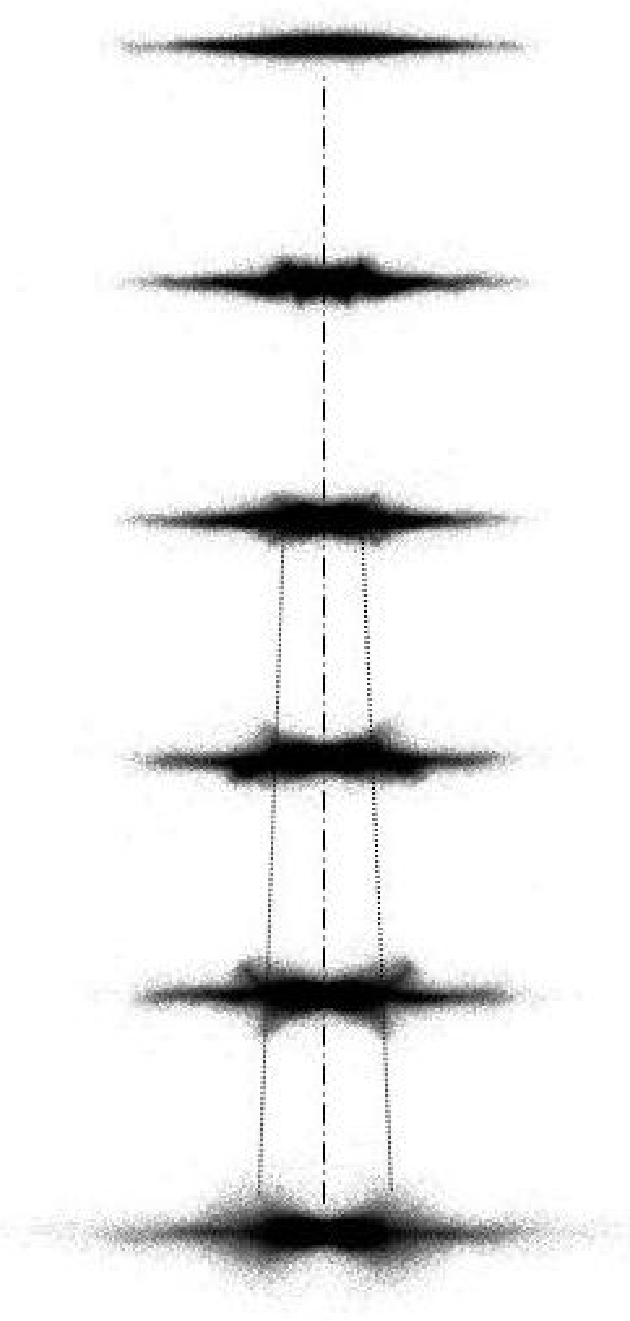}
\includegraphics[width=6.0cm]{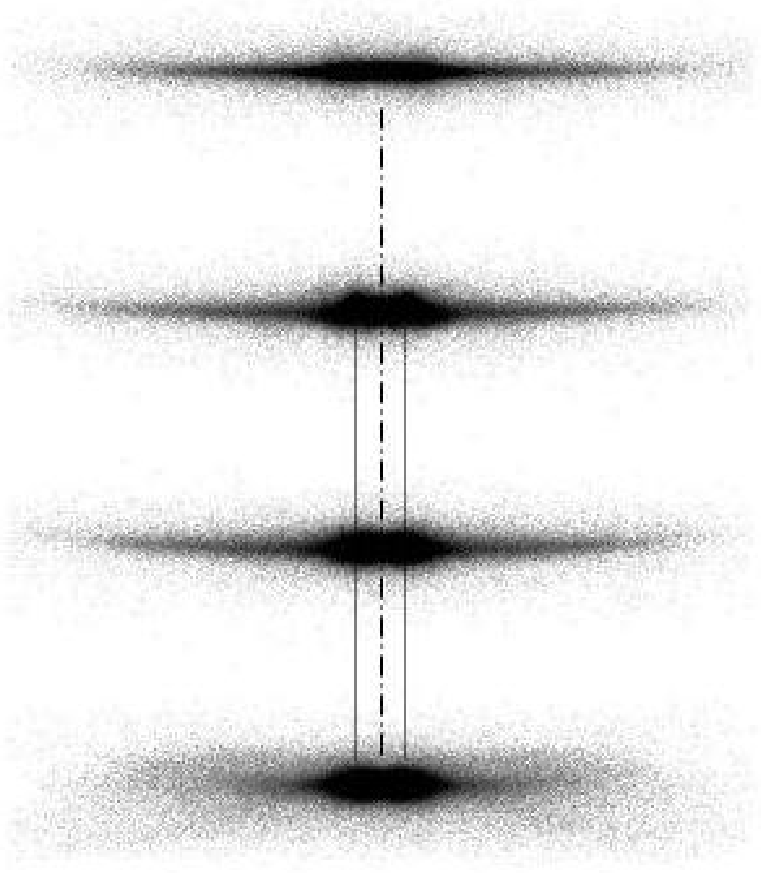}}
\caption{Peanut-shape bulge formation, through vertical resonance with the bar.
With CDM (left), the bar slows down with time, and the resonance moves 
to larger radii. Two peanut features are formed along the evolution, and the last
one is rather extended in radius, while with MOND (right), there is only one peanut
formed, centrally concentrated (Tiret \& Combes 2007). These runs consider only the 
stellar component. Peanuts are less developed, when the disk is rich in gas.}
\label{fig6}
\end{figure}

Summarizing the previous learnings, bars are more frequent in MOND,
and consequently the formation of pseudo-bulges is favored. The fraction
of classical bulges formed in major or minor mergers is likely to be much less,
so that the picture of bulge formation is significantly different in the two regimes.
  These conclusions are applicable mainly to the local galaxies, at very low redshifts.
First bars are less frequent in the past (Sheth et al. 2008), and pseudo-bulges are
thought to be the dominant bulge formation at lower redshift (e.g. Kormendy
\& Kennicutt 2004). Second, it is not well known how the MOND model can be
extended at high redshift. It has been remarked that the critical acceleration
a$_0$ is of the same order as c H$_0$, with H$_0$ the Hubble constant today,
and therefore the critical acceleration could increase with z as H(z). 
Similarly a$_0 \sim$ c ($\Lambda/3)^{1/2}$ (with $\Lambda$ being the dark energy 
parameter), and any kind of variation with time of a$_0$ is possible.

In the standard model, there is another mechanism to form bulges,
which is more dominant at high redshift, that we will consider now.

\begin{figure}[ht]
%\sidecaption
\centerline{
\includegraphics[width=10.5cm]{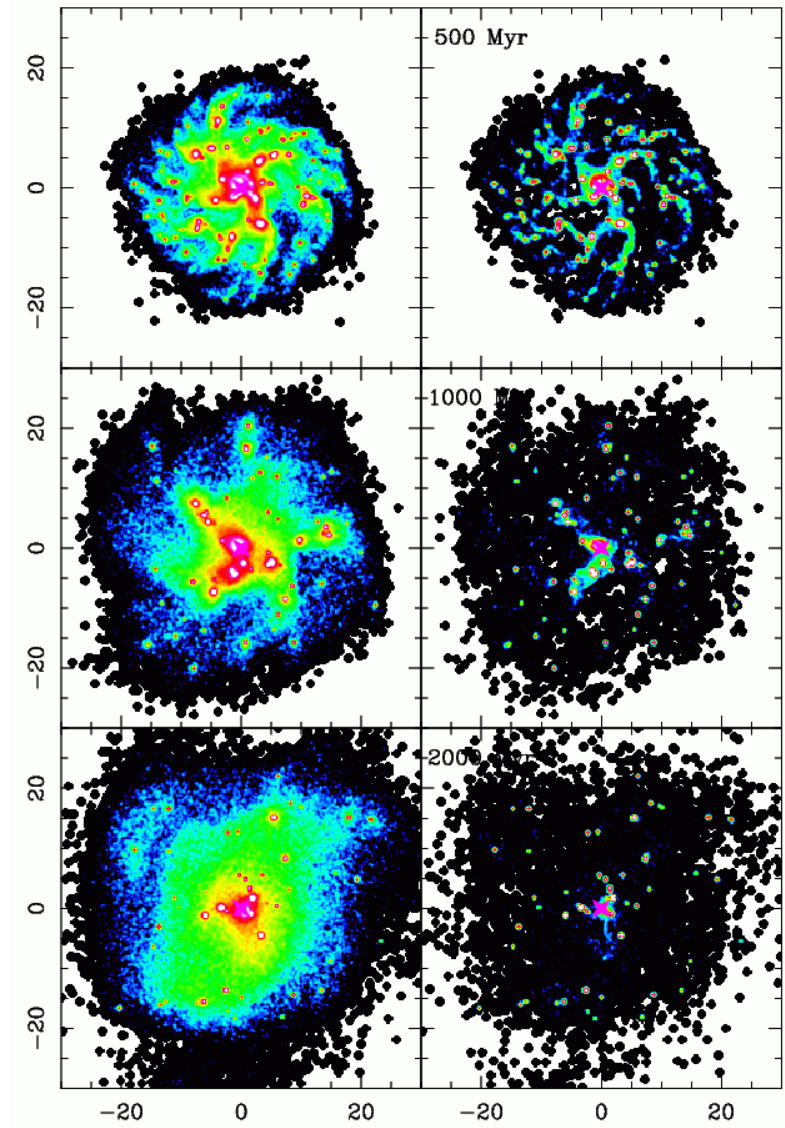}}
\caption{All baryons (left) and gas (right) surface densities of the dwarf
clumpy galaxy, simulated with MOND gravity, at epochs 0.5, 1 and 2 Gyr.
Each panel is 60 kpc in size. The color scale is logarithmic
and the same for all plots. From Combes (2014).}
\label{fig7}
\end{figure}

\section{Clumpy disks}
\label{sec:4}

When the universe was about half of its age (z$\sim$ 0.7) and earlier, the morphology
of spiral galaxies were significantly different from what we know today,
in the Hubble sequence. Galaxies were much more clumpy, with clumps
of gas and stars of kpc size (e.g. Elmegreen 2007).  These very irregular morphologies
are thought to result from the very high gas fraction of these early galaxies.
Noguchi (1999) simulated the formation of galaxies from highly gaseous systems,
and found that they form giant clumps, which by dynamical friction can spiral
inwards to the center rather quickly to form a bulge.  Bournaud et al (2007b)
developed further the dynamical mechanisms, and showed that rather quickly,
clumpy disks form an exponential disk, a bulge, and also a thick disk due to the
stars formed in the turbulently thick gaseous disk. The disruption of the clumps
by the feedback of star formation (supernovae, winds) is not yet well known, 
and can be adjusted to maintain the clumpy disks at the observed frequency
(Elmegreen et al. 2008)
 The large increase of the gas fraction of spiral galaxies with redshift has
been confirmed by direct observations of the molecular gas (e.g. Tacconi et al. 2010).	

\begin{figure}[ht]
%\sidecaption
\centerline{
\includegraphics[angle=-90,width=11.0cm]{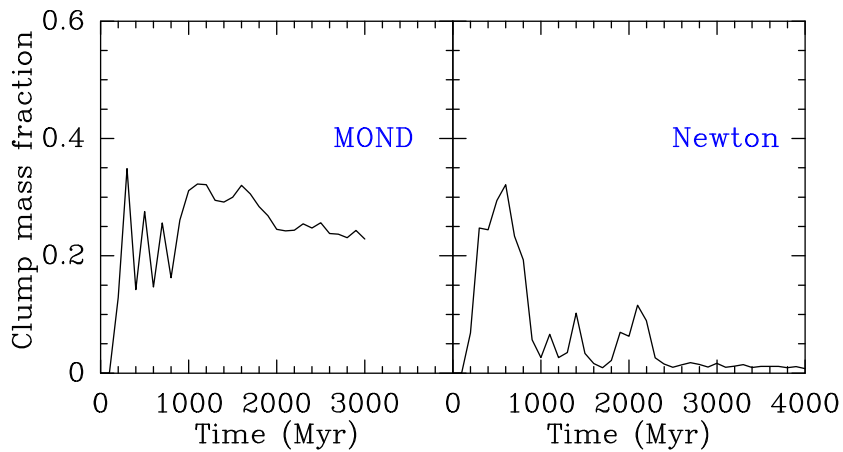}}
\caption{Evolution of the clump mass fraction for
the giant galaxy, in the MOND gravity (left) and
in the Newtonian gravity (right).}
\label{fig8}
\end{figure}

The very high efficiency of bulge formation through dynamical friction in 
clumpy disks might be a problem for the standard dark matter model,
since bulge-less galaxies are quite frequent today (e.g. Weinzirl et al. 2009).
 Since dynamical friction occurs mainly against dark matter halos, it is expected that
it will be much less important in the MOND dynamics, and the rapid
bulge formation could be avoided.
This was indeed demontrated in a recent paper, comparing formation of bulges
in gas-rich clumpy galaxies, in the two gravity models, Newtonian with dark matter and
MOND (Combes, 2014).

This work first computes the dynamical time-scale in an idealized situation,
where the galaxy disks are purely stellar, to isolate the main dynamical phenomenon,
from the more complex gas hydrodynamics, star formation or feedback.
When several clumps are launched randomly in the disk, the dynamical friction
efficiency is difficult to predict, since the wakes of the different massive bodies
interfere (Weinberg 1989). With typical clump mass fraction (25-30\%), in the 
Newtonian model, the 
dynamical time-scale for clumps to spiral into the center of a galaxy with baryonic mass 
6 10$^{10}$ \msol\, 
is 0.3 Gyr, and 1 Gyr for a galaxy with baryonic mass 6 10$^{9}$ \msol. In the MOND regime,
the clumps do not fall into the center before 3 Gyr.
 When the gas and star formation/feedback are taken into account, the simulated galaxy disks
are rapidly unstable to clump formation, due to the gas fraction of 50\%. In
the Newtonian gravity with dark matter, previous results are retrieved, i.e. an increasing
clump mass fraction in the first 200 Myr, and the coalescence of clumps towards the 
center, with a spheroidal bulge formation, in less than 1 Gyr
(Noguchi 1999, Immeli et al 2004, Bournaud et al. 2007b).
With MOND gravity, clumps form quickly too (cf Fig 7), but they maintain in the disk for the 
whole simulation of 3 Gyr, until the gas has been consumed in stars. The clump mass fraction
does not decrease much, being just eroded through stellar feedback, and shear forces (Fig. 8).
 Bulges are clearly not formed in the early clumpy phase of galaxy formation,
as in the Newtonian equivalent systems.

\section{Conclusions}
\label{sec:5}

In the standard model, classical bulges are thought to be formed essentially
in galaxy mergers, which are very frequent in the hierarchical scenario of galaxy
formation. In addition, a small classical bulge is also formed in the first Gyr
of the galaxy lifes, during the clumpy phase, where their disk is gas dominated.
Later on, pseudo-bulges formed out of bar resonances are adding their 
contribution to the classical bulges.

In the frame of MOND, bulges are hardly formed in early times, in the
clumpy phase of galaxy formation, since the dynamical friction without dark matter
halos is not efficient enough to drive clumps towards the center, before they 
are destroyed or reduced by stellar feedback and shear forces.
Classical bulges can form later, through hierarchical merging, with a frequency which is
smaller than what occurs in the analogous Newtonian systems with dark matter.
They however form with comparable frequency through secular evolution,
by vertical resonances with bars.  It is therefore expected that the contribution
of pseudo-bulges with respect to classical bulges is higher in MOND. 
 Globally, bulges are expected less frequent and less massive, which might
be more compatible with observations of local galaxies
(Weinzirl et al. 2009, Kormendy et al. 2010). These tendencies have to be confirmed
with more simulations. 
A complete cosmological context is however not yet possible, given the uncertainties
of the modified gravity models in the early universe.

\bigskip
\parindent=0pt
{\bf References}
\parindent=0pt

%\input{referenc}
%\begin{thebibliography}{}

Athanassoula E.: 2002 ApJ 569, L83
\\Barnes, J. E.: 1988, ApJ 331, 699
\\Barnes, J. E., Hernquist, L. E. 1991, ApJ 370, L65
\\Bekenstein J., Milgrom M.: 1984, ApJ  286, 7
\\Bekenstein J.: 2004 PhRvD 70h3509
\\Bell E.F., Phleps, S., Somerville, R.S. et al.: 2006, ApJ 652, 270
\\Bournaud F., Combes F.: 2002, A\&A 392, 83
\\Bournaud F., Duc P.-A., Masset F.: 2003, A\&A 411, 469
\\Bournaud F., Jog C.J., Combes F.: 2005, A\&A 437, 69
\\Bournaud F., Jog C.J., Combes F.: 2007a, A\&A 476, 1179
\\Bournaud F., Elmegreen B.G., Elmegreen D.M.: 2007b ApJ 670, 237
\\Boylan-Kolchin, M., Bullock, J. S., Kaplinghat, M.: 2011, MNRAS 415, L40
\\Boylan-Kolchin, M., Bullock, J. S., Kaplinghat, M.: 2012, MNRAS 422, 1203
\\Brada R., Milgrom M.: 1999, ApJ 519, 590
\\Bureau, M., Freeman, K. C: 1999, AJ 118, 126
\\Buta R., Combes F.: 1996, F$^{als}$ of Cosmic Physics, Volume 17, pp. 95-281
\\Chandrasekhar S., 1943, ApJ 97, 255
\\Ciotti L., Binney J.: 2004, MNRAS 351, 285
\\Combes F., Sanders R. H.: 1981, A\&A 96, 164
\\Combes F., Debbasch F., Friedli D., Pfenniger D.: 1990, A\&A 233, 82
\\Combes F.: 2014, A\&A, 571, A82
\\de Blok, W. J. G., Walter, F., Brinks, E.. et al.: 2008, AJ 136, 2648
\\Diemand, J., Kuhlen, M., Madau, P. et al.: 2008, Nature 454, 735
\\Di Matteo P., Combes F., Melchior A-L, Semelin B.: 2007, A\&A 468, 61
\\Dubinski J., Mihos J. C., Hernquist L., 1996, ApJ 462, 576
\\Elmegreen D.M.: 2007, in IAU S235, ed. F. Combes \& J. Palous, CUP, p. 376 
\\Elmegreen B.G., Bournaud F., Elmegreen D.M.: 2008, ApJ 688, 67
\\Famaey B., McGaugh S. S.: 2012, Living Reviews in Relativity, vol. 15, no. 10
\\Hibbard J.E., van der Hulst J.M., Barnes J.E., Rich R.M.: 2001, AJ 122, 2969
\\Immeli A., Samland M., Gerhard O., Westera P.: 2004, A\&A  413, 547
\\Kormendy, J., Kennicutt, R. C.: 2004, ARAA 42, 603
\\Kormendy J., Drory N., Bender R., Cornell M. E., 2010, ApJ 723, 54
\\Laurikainen, E., Salo, H., Buta, R., Vasylyev, S.: 2004, MNRAS 355, 1251
\\Laurikainen, E., Salo, H., Buta, R., Knapen, J. H.; 2009, ApJ 692, L34
\\Lopez-Sanjuan, C., Le F\`evre, O., Tasca, L.A.M. et al.: 2013, A\&A 553, A78
\\McGaugh S.S., Schombert J.M., Bothun G.D., de Blok W.J.G.: 2000, ApJ 533, 99
\\Milgrom M.: 1983, ApJ 270, 365
\\Milgrom M.: 1998, ApJ 496, L89 % Galaxy groups
\\Milgrom M., Sanders R.H.: 2007 ApJ 658, 17
\\Moore B., Ghigna S., Governato F. et al.: 1999, ApJ 524, L19
\\Naab, T., Burkert, A. 2003, ApJ 597, 893
\\Navarro, J.F., Steinmetz, M.: 2000, ApJ 528, 607
\\Nipoti, C., Londrillo, P., Ciotti, L., 2007, MNRAS 381, 104
\\Nipoti, C., Ciotti, L., Binney, J., Londrillo, P.: 2008, MNRAS 386, 2194
\\Noguchi M.: 1999, ApJ 514, 77
\\Sanders R.H., McGaugh S.: 2002, ARAA 40, 263
\\Schaye, J., Crain, R. A., Bower, R. G. et al.: 2015, MNRAS 446, 521
\\Sheth, K., Elmegreen, D. M., Elmegreen, B. G. et al.: 2008, ApJ 675, 1141
\\Silk, J., Mamon G.A.: 2012, Research in Astronomy and Astrophysics, Volume 12, Issue 8, pp. 917-946 
\\Stott, J. P., Sobral, D., Smail, I. et al.: 2013, MNRAS 430, 1158
\\Swaters R.A., Sancisi R., van Albada T.S., van der Hulst J.M.: 2009 A\&A 493, 871
\\Tacconi L., Genzel R., Neri R. et al.: 2010, Nature 463, 781
\\Tiret O., Combes F.: 2007, A\&A 464, 517
\\Tiret O., Combes F.: 2008a, A\&A 483, 719
\\Tiret O., Combes F.: 2008b ASPC 396 , 259
\\Toomre A.: 1977, in Evolution of Galaxies and Stellar Populations, Proceedings of a Conference at Yale University, p. 401
\\Vogelsberger, M., Genel, S., Springel, V. et al.: 2014, Nature 509, 177
\\Weinberg, M.D.: 1989, MNRAS 239, 549
\\Weinzirl T., Jogee S., Khochfar S. et al.: 2009, ApJ 696, 411
\\Wu, X., Zhao, H-S., Famaey, B. et al: 2007, ApJ 665, L101
%\end{thebibliography}

\end{document}